\documentclass[aps,preprint]{revtex4}%
\usepackage{amsfonts}
\usepackage{amsmath}
\usepackage{amssymb}
\usepackage{graphicx}%
\providecommand{\U}[1]{\protect\rule{.1in}{.1in}}

\begin{document}
\title[ ]{Understanding Zero-Point Energy in the Context of Classical Electromagnetism}
\author{Timothy H. Boyer}
\affiliation{Department of Physics, City College of the City University of New York, New
York, New York 10031}

\begin{abstract}
Today's textbooks of electromagnetism give the particular solution to
Maxwell's equations involving the integral over the charge and current sources
at retarded times. \ However, the texts fail to emphasize the role played by
the choice of the boundary conditions corresponding to solutions of the
homogeneous Maxwell equations. \ Here we discuss the role of these boundary
conditions for an experimenter with a hypothetical charged harmonic oscillator
as his equipment. \ We describe the observations of the experimenter when
located near a radio station or immersed in thermal radiation at temperature
$T$. \ The classical physicists at the end of the 19th century chose the
homogeneous boundary conditions for Maxwell's equation based upon the
experimental observations of Lummer and Pringsheim which measured only the
thermal radiation which exceeded the random radiation surrounding their
measuring equipment; the physicists concluded that they could take the
homogeneous solutions to vanish at zero temperature. \ Today at the beginning
of the 21st century, classical physicists must choose the homogeneous boundary
conditions for Maxell's equations to correspond to the full radiation spectrum
revealed by the recent Casimir force measurements which detect all the
radiation surrounding conducting parallel plates, including the radiation
absorbed and emitted by the plates themselves. \ The random classical
radiation spectrum revealed by the Casimir force measurements includes
electromagnetic zero-point radiation, which is missing from the spectrum
measured by Lummer and Pringsheim, and which cannot be eliminated by going to
zero temperature. \ This zero-point radiation will lead to zero-point energy
for all systems which have electromagnetic interactions. \ Thus the choice of
the boundary conditions on the homogeneous Maxwell equations is intimately
related to the ideas of zero-point energy and non-radiating ground states
which are introduced in classes of modern physics. \ 

\end{abstract}
\maketitle

\section{Introduction- "The Entire Theoretical Content of Classical
Electrodynamics" \ \ \ \ }

In physics classes today, the subjects of classical electromagnetism and of
modern physics are taught as though there were virtually no connection between
these two areas. Thus electromagnetism calculates the electric and magnetic
fields due to charges and currents, while modern physics introduces the
unfamiliar ideas of zero-point energy and non-radiating ground states. Here we
wish to point out that this continuation of the historical separation is not
something which arises in nature. Some of the ideas of modern physics can be
understood easily and usefully in the context of classical electromagnetism.

When the full Maxwell equations\cite{gaussian}%
\begin{align}
\nabla\cdot\mathbf{E}  &  =4\pi\rho\text{ \ \ \ \ \ \ \ \ \ }\nabla
\cdot\mathbf{B}=0\nonumber\\
\nabla\times\mathbf{E}  &  =-\frac{1}{c}\frac{\partial\mathbf{B}}{\partial
t}\text{ \ \ \ }\nabla\times\mathbf{B}=4\pi\frac{\mathbf{J}}{c}\mathbf{+}%
\frac{1}{c}\frac{\partial\mathbf{E}}{\partial t} \label{M}%
\end{align}
are finally displayed in textbooks of classical electromagnetism, there is
sometimes a generalizing comment on electrodynamics. For example, in Section
7.3.3 of his \textit{Introduction to Classical Electrodynamics},
Griffiths\cite{Griffiths-content} states that Maxwell's equations together
with the Lorentz force law $\mathbf{F}=q[\mathbf{E+(v}/c)\times\mathbf{B]}$
\textquotedblleft summarize the entire theoretical content of classical
electrodynamics.\textquotedblright\ Griffiths goes on to note that some
further information is required regarding matter, and, in a footnote, he
mentions the need for boundary conditions regarding the large-distance
fall-off of electromagnetic fields of bounded charge and current
distributions.\cite{Griffiths-content} \ However, there is no attention paid
to the boundary conditions on the \textit{homogeneous} Maxwell equations where
the charge density $\rho$ and the current density $\mathbf{J}$ are taken to
vanish. The failure to consider the boundary conditions on the homogeneous
Maxwell equations is a serious lapse in electromagnetism textbooks and
represents a continuation of a failure of classical physicists which goes back
to the turn of the 20th century. \ The content of classical electrodynamics
actually includes a crucial choice regarding the boundary conditions on the
solutions of the homogeneous Maxwell's equations. \ In the discussion to
follow, we will try to make the important nature of this boundary condition
clear to students of classical electrodynamics, and will show its connection
to ideas of zero-point energy. \ 

\section{Plane Waves and Charged Harmonic Oscillators}

\subsubsection{Plane Waves as Solutions of the Homogeneous Maxwell Equations}

In order to connect classical electrodynamics with the ideas of zero-point
energy and non-radiating ground states, it is useful to consider plane waves
and charged harmonic oscillator systems within classical electrodynamics.
\ Plane waves are solutions of the homogeneous Maxwell equations (\ref{M})
where $\rho$ and $\mathbf{J}$ are taken to vanish. \ All textbooks of
electromagnetism introduce plane waves\cite{Griffiths-pw}%
\begin{align}
\mathbf{E(r,}t)  &  =\widehat{\epsilon}E_{0}\cos(\mathbf{k}\cdot
\mathbf{r}-\omega t+\theta)\nonumber\\
\mathbf{B(r,}t)  &  =\widehat{k}\times\widehat{\epsilon}E_{0}\cos
(\mathbf{k}\cdot\mathbf{r}-\omega t+\theta) \label{PW}%
\end{align}
where $\widehat{\epsilon}$ is a unit vector orthogonal to the wave vector
$\mathbf{k}$, $E_{0}$ sets the scale for the wave, $\omega=ck$ is the angular
frequency, and $\theta$ provides a determination of phase. \ The introduction
of plane waves means precisely making a choice regarding the boundary
conditions on the homogeneous Maxwell equations. \ 

\subsubsection{Use of the Scalar and Vector Potentials}

Now when discussing the sources of radiation in classical electrodynamics, it
is convenient to calculate the scalar and vector potentials $\Phi$
and~$\mathbf{A}$, which are connected to the fields by $\mathbf{E}=-\nabla
\Phi-(1/c)\partial\mathbf{A}/\partial t$ and $\mathbf{B=\nabla\times A}.$
Maxwell's equations in terms of the potentials, written in the Lorentz (or
Lorenz) gauge, take the form\cite{Griffiths-pots}%
\begin{align}
\left(  \nabla^{2}-\frac{1}{c^{2}}\frac{\partial^{2}}{\partial t^{2}}\right)
\Phi &  =-4\pi\rho\nonumber\\
\left(  \nabla^{2}-\frac{1}{c^{2}}\frac{\partial^{2}}{\partial t^{2}}\right)
\mathbf{A}  &  =-4\pi\frac{\mathbf{J}}{c} \label{POTs}%
\end{align}
The scalar and vector potentials for the plane wave in Eq. (\ref{PW}) can be
written as%
\begin{align}
\Phi(\mathbf{r,}t)  &  =0\nonumber\\
\mathbf{A(r},t)  &  =\frac{c}{\omega}\widehat{\epsilon}E_{0}\sin
(\mathbf{k}\cdot\mathbf{r}-\omega t+\theta) \label{A}%
\end{align}
These plane-wave potentials in Eq. (\ref{A}) provide a solution of the
homogeneous Maxwell equations for the potentials in Eq. (\ref{POTs}).

\subsubsection{General Solution of Maxwell's Equations}

In terms of the scalar and vector potentials, the general solution of
Maxwell's equations takes the form
\begin{align}
\Phi(\mathbf{r,}t)  &  =\Phi^{in}(\mathbf{r},t)+%
{\textstyle\int}
d^{3}r^{\prime}%
{\textstyle\int}
dt^{\prime}\frac{\delta(t-t^{\prime}-|\mathbf{r-r}^{\prime}|/c)}%
{|\mathbf{r-r}^{\prime}|}\rho(\mathbf{r}^{\prime},t^{\prime})\nonumber\\
&  =\Phi^{in}(\mathbf{r},t)+%
{\textstyle\int}
d^{3}r^{\prime}\frac{\rho(\mathbf{r,}t-|\mathbf{r-r}^{\prime}|/c)}%
{|\mathbf{r-r}^{\prime}|} \label{GS1}%
\end{align}%
\begin{align}
\mathbf{A}(\mathbf{r,}t)  &  =\mathbf{A}^{in}(\mathbf{r},t)+%
{\textstyle\int}
d^{3}r^{\prime}%
{\textstyle\int}
dt^{\prime}\frac{\delta(t-t^{\prime}-|\mathbf{r-r}^{\prime}|/c)}%
{|\mathbf{r-r}^{\prime}|}\frac{\mathbf{J}(\mathbf{r}^{\prime},t^{\prime})}%
{c}\nonumber\\
&  =\mathbf{A}^{in}(\mathbf{r},t)+%
{\textstyle\int}
d^{3}r^{\prime}%
{\textstyle\int}
dt^{\prime}\frac{\mathbf{J}(\mathbf{r,}t-|\mathbf{r-r}^{\prime}|/c)}%
{c|\mathbf{r-r}^{\prime}|} \label{GS2}%
\end{align}
where $\Phi^{in}(\mathbf{r},t)$ and $\mathbf{A}^{in}(\mathbf{r},t)$ are
solutions of the homogeneous Maxwell's equations, and the integrals provide
the fields due to the sources at the retarded time. \ If the electrodynamic
system included plane waves as in Eq. (\ref{A}), then these would appear as
$\Phi^{in}(\mathbf{r},t)$ and $\mathbf{A}^{in}(\mathbf{r},t).$ \ 

The general solution of a system of linear differential equations is always
written in the form of Eqs. (\ref{GS1}) and (\ref{GS2}) which includes both
the general solution of the homogeneous differential equations and a
particular solution of the inhomogeneous differential equations. \ However,
for some reason, the textbooks of classical electromagnetism never write the
general solution of Maxwell's equations in this correct form.\cite{Coleman}%
\ For some reason, the homogeneous solution part involving $\Phi
^{in}(\mathbf{r},t)$ and $\mathbf{A}^{in}(\mathbf{r},t)$ is always omitted in
textbooks, leaving only the integral over the sources multiplying the retarded
Green function for the scalar wave equation in all spacetime.\cite{texts}

\subsubsection{Charged Harmonic Oscillator System}

One of the simplest systems used in classical electrodynamics is the charged
harmonic oscillator system, which can be pictured as a particle of charge $e$
and mass $m$ at one end of a spring of spring-constant $\kappa$ (with the
other end attached to a wall), so that (neglecting the effects of damping) the
oscillator has natural frequency of oscillation $\omega_{0}=(\kappa/m)^{1/2}.$
\ \ Such charged harmonic oscillator systems are often used in discussions of
dispersion and absorption in electromagnetism texts.\cite{Griffiths-disp}.
\ The equation of motion for a small charged harmonic oscillator
system\ located on the $x$-axis with equilibrium position at the origin of
coordinates is\cite{Griffiths-osc}%
\begin{equation}
m\frac{d^{2}x}{dt^{2}}=-m\omega_{0}^{2}x-m\gamma\frac{dx}{dt}+eE_{x}(0,t)
\label{SHO}%
\end{equation}
where $-m\gamma dx/dt$ is the damping force which is proportional to the
velocity, and $eE_{x}(0,t)$ represents the force applied to the oscillator by
the electric field which exists in the region.\ \ If the electric field
driving the oscillator corresponds to the plane wave given in Eqs. (\ref{PW})
and (\ref{A}), then the steady-state solution for the displacement of the
oscillator is\cite{Griffiths-osc2}
\begin{equation}
x(t)=\operatorname{Re}\frac{e\epsilon_{x}E_{0}\exp[i(-\omega t+\theta
)]}{m(-\omega^{2}+\omega_{0}^{2}+i\omega\gamma)} \label{X}%
\end{equation}
We notice that the oscillator has an oscillating position whose time-average
is zero $\left\langle x\right\rangle =0,$ but whose mean square is given by
\begin{equation}
\left\langle x^{2}\right\rangle =\frac{e^{2}\epsilon_{x}^{2}E_{0}^{2}}%
{2m^{2}[(-\omega^{2}+\omega_{0}^{2})^{2}+(\omega\gamma)^{2}]} \label{X2}%
\end{equation}
If we take the time derivative in Eq.(\ref{X}) to obtain $\dot{x}(t),$ then
the velocity is oscillating with an average which vanishes $\left\langle
\dot{x}\right\rangle =0$ and a mean square given by%
\begin{equation}
\left\langle \dot{x}^{2}\right\rangle =\frac{e^{2}\omega^{2}\epsilon_{x}%
^{2}E_{0}^{2}}{2m^{2}[(-\omega^{2}+\omega_{0}^{2})^{2}+(\omega\gamma)^{2}]}
\label{Xdot}%
\end{equation}
The energy of the oscillator is accordingly%
\begin{equation}
\left\langle \mathcal{E}\right\rangle =\left\langle \frac{1}{2}m\dot{x}%
^{2}+\frac{1}{2}m\omega_{0}^{2}x^{2}\right\rangle =\frac{(\omega^{2}%
+\omega_{0}^{2})e^{2}\epsilon_{x}^{2}E_{0}^{2}}{4m[(-\omega^{2}+\omega_{0}%
^{2})^{2}+(\omega\gamma)^{2}]} \label{En}%
\end{equation}

The oscillator acts as a small electric dipole oscillator with electric dipole
moment $\mathbf{p}(t)=\widehat{i}ex(t),$ whose charge density and current
density can be approximated as\cite{Zangwill-dip}%
\begin{equation}
\rho(\mathbf{r},t)=-\nabla\cdot\lbrack\mathbf{p}(t)\delta^{3}(\mathbf{r)]}
\label{R}%
\end{equation}%
\begin{equation}
\mathbf{J(r},t)=\frac{d\mathbf{p}}{dt}\delta^{3}(\mathbf{r)} \label{J}%
\end{equation}
The charge and current distributions are sources for electromagnetic fields as
given by Eqs. (\ref{GS1}) and (\ref{GS2}). \ In the context of Maxwell's
equations for the electromagnetic fields of our system, the plane wave
appearing in Eqs. (\ref{PW}) and (\ref{A}) and also appearing as the driving
force in Eq. (\ref{SHO}) is a homogeneous solution of Maxwell's equations in
Eqs. (\ref{GS1}) and (\ref{GS2}) while the radiation emitted by the oscillator
depends on the charge and current distributions in Eqs.(\ref{R}) and
(\ref{J}), giving%
\begin{align}
\Phi(\mathbf{r,}t)  &  =0+%
{\textstyle\int}
d^{3}r^{\prime}%
{\textstyle\int}
dt^{\prime}\frac{\delta(t-t^{\prime}-|\mathbf{r-r}^{\prime}|/c)}%
{|\mathbf{r-r}^{\prime}|}\left\{  -\nabla^{\prime}\cdot\lbrack\widehat{i}%
ex(t^{\prime})\delta^{3}(\mathbf{r}^{\prime}\mathbf{)]}\right\} \nonumber\\
&  =-\nabla\cdot\lbrack\widehat{i}ex(t-r/c)] \label{VSol}%
\end{align}%
\begin{align}
\mathbf{A}(\mathbf{r,}t)  &  =\frac{c}{\omega}\widehat{\epsilon}E_{0}%
\sin(\mathbf{k}\cdot\mathbf{r}-\omega t+\theta)+%
{\textstyle\int}
d^{3}r^{\prime}%
{\textstyle\int}
dt^{\prime}\frac{\delta(t-t^{\prime}-|\mathbf{r-r}^{\prime}|/c)}%
{|\mathbf{r-r}^{\prime}|}\left\{  \frac{d\mathbf{p}}{dt^{\prime}}\frac
{\delta^{3}(\mathbf{r}^{\prime}\mathbf{)}}{c}\right\} \nonumber\\
&  =\frac{c}{\omega}\widehat{\epsilon}E_{0}\sin(\mathbf{k}\cdot\mathbf{r}%
-\omega t+\theta)+\mathbf{\dot{p}(}t-r/c)/c \label{ASol}%
\end{align}
Here the potentials $\Phi(\mathbf{r,}t)$ and $\mathbf{A}(\mathbf{r,}t)$ allow
us to obtain the full electromagnetic fields associated with the system
consisting of the plane wave and the oscillator.

It must be emphasized that the charged harmonic oscillator located in the
plane wave is both absorbing energy from the incident plane wave and is also
losing energy to emitted radiation. \ The energy loss and energy gain are in
balance when the oscillator has the energy given in Eq. (\ref{En}). \ The
oscillator does not radiate away all its energy and collapse to zero energy
because of the presence of the driving force of the plane wave corresponding
to a solution of the homogeneous Maxwell's equations.

\section{Understanding Boundary Conditions on the Homogeneous Maxwell
Equations in Nature}

\subsubsection{An Experimenter with a Charged Harmonic Oscillator}

Having introduced plane waves as examples of solutions of the homogeneous
Maxwell equations and charged harmonic oscillators as simple classical
electrodynamic systems, we wish to explore the role of the homogeneous
boundary conditions on Maxwell's equations in nature. Let us imagine an
experimenter with a charged harmonic oscillator in his laboratory. As far as
the experimenter is concerned, the homogeneous solutions of Maxwell's
equations $\Phi^{in}$ and $\mathbf{A}^{in}$ are the fields which exist before
he turns on his equipment and which are not at his control. \ If an
experimenter is located in a place where the only electromagnetic fields are
those which are produced by his own equipment, then he takes the homogeneous
solutions of Maxwell's equations to vanish, and he can arrange to start his
experiments with his harmonic oscillator at rest at his origin of coordinates
$x(t)=0$, $\dot{x}(t)=0$. However, if the experimenter is in a location where
a nearby radio station is emitting radiation, then the experimenter will find
that his charged oscillator has a non-zero amplitude due to forcing by the
electromagnetic fields of the radio station. If the experimenter is confined
to a small laboratory compared to the distance to the radio station, the
experimenter would describe the electromagnetic fields of the radio station as
plane waves satisfying the homogeneous Maxwell equations while the radiation
emitted by his oscillator would give additional radiation. \ His oscillator is
responding to the homogeneous solution which is not under his direct control.
\ Indeed, the physical situation corresponds exactly to the electromagnetic
fields following from the potentials given in Eqs. (\ref{VSol}) and
(\ref{ASol}). If the radio station turned off for the night, then the
\ homogeneous fields $\Phi^{in}$ and $\mathbf{A}^{in}$ for the experimenter
would drop to zero, and the oscillator would emit all its energy as radiation
and would collapse back to $x(t)=0$, $\dot{x}(t)=0$.

Of course, the need for the experimenter to include the solutions of the
homogeneous Maxwell equations $\Phi^{in}$ and $\mathbf{A}^{in}$ will depend
upon the relative magnitudes of the energy delivered to his oscillator by his
own equipment compared to the energy delivered to his oscillator by the
homogeneous solution of Maxwell's equations. \ If the energy delivered to the
oscillator by his own equipment is overwhelmingly larger than the energy
delivered by the homogeneous solution, then he can ignore the contribution of
the homogenous solution. \ This latter situation seems to be that envisioned
by the textbooks of classical electromagnetism which omit entirely the terms
$\Phi^{in}$ and $\mathbf{A}^{in}$ involving the homogeneous solution.

Now the resonant circuits measured by students in our elementary physics
courses and indeed our modern radio receivers act like little harmonic
oscillators with natural resonant frequencies $\omega_{0}$, analogous to the
harmonic oscillator of our imagined experimenter. \ The energy delivered by a
student-lab signal generator is overwhelmingly larger than that present in the
ambient radiation in the laboratory, and so the effects of the homogeneous
solution are not discussed. \ However, in our elementary labs, one sometimes
asks students to touch the input lead to an oscilloscope to show that the
laboratory classroom is actually filled with electromagnetic waves (usually
generated by the classroom wiring or the fluorescent lamps) which would be
described as solutions $\Phi^{in}$ and $\mathbf{A}^{in}$ of the homogeneous
Maxwell equations which are not under the control of the equipment on the
laboratory table. \ 

\subsubsection{An Experimenter in Thermal Radiation}

Suppose now that the experimenter with his harmonic oscillator is interested
in making delicate measurements of very low-energy phenomena. \ We imagine
that he is located not near a radio station but rather in a room at non-zero
temperature $T,$ so that the solution of the homogeneous Maxwell equations
corresponds to thermal electromagnetic radiation. Within classical physics,
thermal radiation is described as random electromagnetic radiation with a
characteristic \textquotedblleft thermal\textquotedblright\ spectrum
corresponding to the spectrum of blackbody radiation. \ Taking periodic
boundary conditions for a very large box of volume $V$, we can treat thermal
radiation as a sum over plane waves of all wave vectors. \ Following the form
in Eq. (\ref{A}), we can choose the scalar potential $\Phi_{T}^{in}%
(\mathbf{r},t)=0$ and the vector potential as%
\begin{equation}
\mathbf{A}_{T}^{in}\mathbf{(r},t)=%
{\textstyle\sum_{\mathbf{k}}}
{\textstyle\sum_{\lambda=1}^{2}}
\frac{c}{\omega}\widehat{\epsilon}(\mathbf{k},\lambda)\left(  \frac{8\pi
U(\omega,T)}{V}\right)  ^{1/2}\left\{  \sin[\mathbf{k\cdot r}-\omega
t+\theta(\mathbf{k},\lambda)]\right\}  \label{Atherm}%
\end{equation}
where the wave vectors $\mathbf{k}$ correspond to $\mathbf{k=}\widehat{i}%
l2\pi/a+\widehat{j}m2\pi/a+\widehat{k}n2\pi/a$ with $l,m,n$ running over all
positive and negative integers, $a$ is a length such that $a^{3}=V,$ and the
two mutually-orthogonal polarization vectors $\widehat{\epsilon}%
(\mathbf{k},\lambda)$ are orthogonal to the wave vectors $\mathbf{k.}$ Since
thermal radiation is isotropic in the inertial frame of its container, the
amplitude $[U(\omega,T)]^{1/2}$ depends only on the frequency $\omega
=c|\mathbf{k}|=ck,$ and the constants are chosen so that $U(\omega,T)$\ is the
energy per normal mode appropriate for the thermal radiation spectrum in
classical physics. \ In order to describe the randomness of the radiation, the
phases $\theta(\mathbf{k},\lambda)$ are chosen as random variables uniformly
distributed on $(0,2\pi],$ independently distributed for each $\mathbf{k}$ and
$\lambda.$

The electric and magnetic fields of the thermal radiation in the
experimenter's laboratory would set his dipole oscillator into oscillation,
just as the electric and magnetic fields of the radio station set the
oscillator into oscillation according to Eq. (\ref{SHO}). \ For thermal
radiation as given in Eq. (\ref{Atherm}) and the oscillator located at the
coordinate origin, the steady-state solution corresponding to Eq.(\ref{X})
becomes%
\begin{equation}
x(t)=\operatorname{Re}%
{\textstyle\sum_{\mathbf{k}}}
{\textstyle\sum_{\lambda=1}^{2}}
\epsilon_{x}(\mathbf{k},\lambda)\left(  \frac{8\pi U(\omega,T)}{V}\right)
^{1/2}\frac{e\exp\left\{  i[-\omega t+\theta(\mathbf{k},\lambda)]\right\}
}{m(-\omega^{2}+\omega_{0}^{2}+i\omega\gamma)} \label{Xt}%
\end{equation}
Again, the experimenter would describe the forcing electromagnetic fields
acting on his oscillator as solutions of the homogeneous Maxwell's equations
while the electromagnetic fields radiated by his oscillator would be given by
the retarded fields associated with the charges densities and current
densities of the oscillator motion, just as indicated in Eqs. (\ref{GS1}) and
(\ref{GS2}).

In thermal radiation, the mean displacement of the oscillator and the mean
velocity are both zero $\left\langle x(t)\right\rangle =0,$ $\left\langle
\dot{x}(t)\right\rangle =0,$ but the mean squares are non-zero. \ We can find
the mean-square displacement by averaging over time or averaging over the
random phases at a fixed time. \ Since the random phases $\theta
(\mathbf{k},\lambda)$ are distributed randomly and independently for each
mode, we have the averages
\begin{equation}
\left\langle \exp\left\{  i[-\omega t+\theta(\mathbf{k},\lambda)]\right\}
\exp\left\{  i[-\omega^{\prime}t+\theta(\mathbf{k}^{\prime},\lambda^{\prime
})]\right\}  \right\rangle =0 \label{av1}%
\end{equation}
and
\begin{equation}
\left\langle \exp\left\{  i[-\omega t+\theta(\mathbf{k},\lambda)]\right\}
\exp\left\{  -i[-\omega^{\prime}t+\theta(\mathbf{k}^{\prime},\lambda^{\prime
})]\right\}  \right\rangle =\delta_{\mathbf{kk}^{\prime}}\delta_{\lambda
\lambda^{\prime}} \label{av2}%
\end{equation}
which give%
\begin{equation}
\left\langle x^{2}\right\rangle =%
{\textstyle\sum_{\mathbf{k}}}
{\textstyle\sum_{\lambda=1}^{2}}
\epsilon_{x}^{2}(\mathbf{k},\lambda)\left(  \frac{8\pi U(\omega,T)}{V}\right)
\frac{e^{2}}{2m^{2}[(-\omega^{2}+\omega_{0}^{2})^{2}+(\omega\gamma)^{2}]},
\label{X2av}%
\end{equation}%
\begin{equation}
\left\langle \dot{x}^{2}\right\rangle =%
{\textstyle\sum_{\mathbf{k}}}
{\textstyle\sum_{\lambda=1}^{2}}
\epsilon_{x}^{2}(\mathbf{k},\lambda)\left(  \frac{8\pi U(\omega,T)}{V}\right)
\frac{e^{2}\omega^{2}}{2m^{2}[(-\omega^{2}+\omega_{0}^{2})^{2}+(\omega
\gamma)^{2}]}, \label{Xdot2av}%
\end{equation}
and the average energy of the oscillator
\begin{equation}
\left\langle \mathcal{E}(\omega_{0},T\right\rangle =%
{\textstyle\sum_{\mathbf{k}}}
{\textstyle\sum_{\lambda=1}^{2}}
\epsilon_{x}^{2}(\mathbf{k},\lambda)\left(  \frac{8\pi U(\omega,T)}{V}\right)
\frac{e^{2}(\omega^{2}+\omega_{0}^{2})}{4m[(-\omega^{2}+\omega_{0}^{2}%
)^{2}+(\omega\gamma)^{2}]} \label{Eav}%
\end{equation}
Now we are imagining that the box (or laboratory) which holds the thermal
radiation is sufficiently large that the normal modes are very closely spaced,
and therefore the sum over normal modes can be replaced by an integral,%
\begin{equation}
\left\langle \mathcal{E}(\omega_{0},T)\right\rangle =\left(  \frac{a}{2\pi
}\right)  ^{3}\int d^{3}k%
{\textstyle\sum_{\lambda=1}^{2}}
\epsilon_{x}^{2}(\mathbf{k},\lambda)\left(  \frac{8\pi U(\omega,T)}{V}\right)
\frac{e^{2}(\omega^{2}+\omega_{0}^{2})}{4m[(-\omega^{2}+\omega_{0}^{2}%
)^{2}+(\omega\gamma)^{2}]} \label{Eav2}%
\end{equation}
which is sharply peaked at $\omega=\omega_{0}.$

Although the textbooks of classical electrodynamics are often interested in a
damping force corresponding to the absorption of energy by the material
forming the oscillator, we are interested in the simplest possible case where
the damping term represents an approximation to the \textit{radiation} damping
force $F_{rr}=[(2e^{2})/(3c^{3})]d^{3}x/dt^{3}\approx-\omega_{0}^{2}%
[(2e^{2})/(3c^{3})]dx/dt,$ so that $\gamma=\omega_{0}^{2}(2e^{2})/(3mc^{3}).$
\ In this case, we can integrate over all angles so that $\epsilon_{x}%
^{2}(\mathbf{k},\lambda)$ contributes a factor of 1/3 for each polarization,
and then approximate the integral over $k=\omega/c$ by extending the lower
limit to minus infinity, setting $\omega=\omega_{0}$ in every term except for
$(-\omega^{2}+\omega_{0}^{2})\approx2\omega_{0}(\omega_{0}-\omega),$ and using
the definite integral%
\begin{equation}%
{\textstyle\int_{-\infty}^{\infty}}
\frac{dx}{a^{2}x^{2}+b^{2}}=\frac{\pi}{ab} \label{I}%
\end{equation}
to obtain for the average oscillator energy\cite{Lavenda}%
\begin{equation}
\left\langle \mathcal{E}(\omega_{0},T)\right\rangle =U(\omega_{0},T)
\label{Eav3}%
\end{equation}
Thus the average energy $\left\langle \mathcal{E}(\omega_{0},T)\right\rangle $
of the oscillator with resonant frequency $\omega_{0}$ is the same as the
average energy $U(\omega_{0},T)$ of the radiation normal mode at the same frequency.

\subsubsection{Random Classical Radiation Known at the End of the 19th
Century}

The description we have given for the classical electrodynamics involving a
charged harmonic oscillator corresponds exactly to that involved in the
thinking of the physicist at the end of the 19th century. \ It parallels the
classical development given by Planck in his theory of thermal
radiation.\cite{Planck} \ Now the experiments of Lummer and
Pringsheim\cite{LP} in 1899 were fitted by Planck\cite{Planck2} in 1900 to a
spectrum for thermal radiation%
\begin{equation}
U_{P}(\omega,T)=\hbar\omega\frac{1}{\exp[\hbar\omega/k_{B}T]-1}=\frac{1}%
{2}\hbar\omega\coth\left(  \frac{\hbar\omega}{2k_{B}T}\right)  -\frac{1}%
{2}\hbar\omega\label{LP}%
\end{equation}
which is now termed the Planck spectrum. \ For a non-zero value of temperature
$T$, the Planck radiation spectrum given in Eq. (\ref{LP}) goes to zero for
large values of frequency $\omega$ because of the exponential function
appearing in the denominator$.$ \ At low frequencies $\omega$, the spectrum in
Eq. (\ref{LP}) goes over to
\begin{equation}
U_{P}(\omega,T)\approx k_{B}T-\frac{1}{2}\hbar\omega+O(\hbar\omega
/k_{B}T)\text{ \ \ \ for \ \ }k_{B}T>>\hbar\omega\label{LP2}%
\end{equation}
The low-frequency value $U_{P}(\omega,T)\approx k_{B}T$ fits with the
equipartition result of non-relativistic classical statistical mechanics for a
harmonic oscillator at temperature $T.$

Now an experimenter in a laboratory at room temperature would have a difficult
time measuring the thermal radiation of his harmonic oscillator. At room
temperature, the energy per normal mode gives $k_{B}T=1/40eV.$ \ This energy
is vastly smaller than the energies of the macroscopic harmonic oscillators of
our elementary mechanics courses or of the resonant circuits in our elementary
electromagnetism classes. \ Rather, this value corresponds to energy at the
atomic scale. \ We have no macroscopic evidence regarding a single oscillator
at this scale. \ On the other hand,\ if we go to high enough temperatures for
a single oscillator to have a macroscopic energy, the thermal energy would
destroy the harmonic oscillator system. \ Lummer and Pringsheim's experiments
escaped the limitations of a single oscillator by measuring the thermal energy
received by a macroscopic surface, and so involved an enormous number of oscillators.\ \ \ 

\subsubsection{Random Classical Radiation Known at the Beginning of the 21th
Century}

When continuing our story of the experimenter with his harmonic oscillator in
his laboratory, we will want to use not only the experimental information
available at the end of the 19th century but also the experimental information
available at the beginning of the 21st century. It turns out that the thermal
experiments of 1899 contained a crucial limitation which confused the
classical electromagnetic theorists at the beginning of the 20th century.
\ The Lummer-Pringsheim experiments of 1899 measured only the random radiation
of their sources which was above the random radiation surrounding their
measuring devices. If their sources were at the same temperature as their
measuring devices, the measuring devices would have registered no signal at
all. Today, in contrast to the end of the 19th century, random classical
radiation measurements are available which are of an entirely different
character from those of Lummer and Pringsheim.

Textbooks of classical electromagnetism point out that radiation falling on a
conducting surface will place a force on the surface.\cite{Griffiths-press}
Today it is possible to make measurements of the forces, termed Casimir
forces,\cite{Casimir} between parallel conducting plates due to any radiation
which is present surrounding the plates. These force measurements allow the
measurement of all the radiation falling on the plates. In contrast to the
limitation present in the thermal measurements at the end of the 19th century,
any random radiation emitted, absorbed or reflected by the plates themselves
contributes to the forces between the plates and will be registered by force
measurements on the plates. \ The forces may be small, but since the area of
the plates can be made large, the forces are measurable as macroscopic forces.
\ For plates of area 1 cm$^{2}$, and separations of half a micron, the force
is measured as 1/4 dyne.

In the measurements of Casimir forces, it is the wavelengths roughly
comparable to the plate separation which give the dominant force contribution.
\ Thus at large parallel-plate separations, the long wavelength radiation
corresponding to the Rayleigh-Jeans spectrum contributes. \ At small
separations, the short-wavelength high-frequency waves provide the dominant
force, and, according to the Planck expression (\ref{LP}) corresponding to
Lummer and Pringsheim's thermal measurements, the radiation goes to zero at
low temperatures. \ However, when the Casimir-force measurements between
parallel conducting plates are actually carried out at low temperature, it is
found that the forces between the plates do not correspond to the spectrum of
Lummer and Pringsheim's thermal measurements. Rather the forces between the
plates do not go to zero as the temperature is decreased.\cite{Casimir2} Thus
instead of the spectrum given in Eq. (\ref{LP}), the spectrum of random
electromagnetic radiation found from modern Casimir force experiments contains
an additional part at low temperatures. \ The spectrum of random
electromagnetic radiation which fits both the thermal measurements of Lummer
and Pringsheim and also the Casimir force measurements corresponds to an
energy per normal mode%
\begin{equation}
U_{C}(\omega,T)=\frac{1}{2}\hbar\omega\coth\left(  \frac{\hbar\omega}{2k_{B}%
T}\right)  =\hbar\omega\frac{1}{\exp[\hbar\omega/k_{B}T]-1}+\frac{1}{2}%
\hbar\omega\label{Cf}%
\end{equation}
At high temperatures, this spectrum again goes over to the equipartition
result
\begin{equation}
U_{C}(\omega,T)\approx k_{B}T+O(\hbar\omega/k_{B}T)\text{ \ \ for~}%
k_{B}T>>\hbar\omega\label{RJ2}%
\end{equation}
However, at low temperatures, the spectrum has an energy per normal mode%
\begin{equation}
U_{C}(\omega,T)\approx(1/2)\hbar\omega+O[\exp(-\hbar\omega/k_{B}T)]\text{
\ \ for~}k_{B}T<<\hbar\omega\label{ZPR}%
\end{equation}
corresponding to a zero-point spectral energy per normal mode $U_{C}%
(\omega,0)=U_{ZP}(\omega)$ with
\begin{equation}
U_{ZP}(\omega)=(1/2)\hbar\omega\label{ZPR2}%
\end{equation}

Let us now return to our hypothetical experimenter with his charged harmonic
oscillator. \ Since we have no direct macroscopic measurements of a single
harmonic oscillator in thermal radiation, we must infer the behavior of the
oscillator of our imagined experimenter. \ Clearly, we wish to use the best
possible experimental information in order to infer this behavior. \ The
latest experimental information indicates that the random radiation which
would influence the oscillator of our imagined experimenter in his laboratory
does not correspond to the 1899 thermal spectrum in Eq. (\ref{LP}) but rather
to the full spectrum of Eq. (\ref{Cf}).

\subsubsection{Classical Electromagnetic Zero-Point Radiation}

The radiation energy per normal mode given in Eq. (\ref{Cf}) corresponds
closely to the spectrum found in the 1899 measurements of Lummer and
Pringsheim, but it also includes additional random radiation which exists even
as the temperature $T$ goes to zero. This random radiation which exists at
zero-temperature is termed zero-point radiation. Since here we are describing
nature with a purely classical electromagnetic theory, we will term this
radiation "classical electromagnetic zero-point radiation." \ 

Based on the 1899 thermal spectrum in Eq. (\ref{LP}), the classical physicists
at the beginning of the 20th century believed that they could take the
solutions of the homogeneous Maxwell equations to vanish by screening their
laboratories and going to the absolute zero of temperature. \ Lorentz makes
this explicit in his monograph \textit{The Theory of Electrons}.\cite{Lorentz}
\ Today at the beginning of the 21st century, a classical physicist cannot
make this assumption. \ Our imagined experimenter in his laboratory may cool
his laboratory close to absolute zero, yet he can not escape the random
zero-point radiation revealed by experimental measurements of Casimir forces.
\ Any classical description of electromagnetism in accord with nature must
include zero-point radiation which corresponds to the solution of the
homogeneous Maxwell equations relevant at $T=0$. \ 

\subsubsection{Classical Zero-Point Energy and Classical Ground States}

It must be emphasized that the classical electromagnetic zero-point radiation
appearing in the experiments on Casimir forces will also influence the charged
harmonic oscillator of our experimenter in his laboratory. If the experimenter
is far from any radio station and he also cools his laboratory to near
absolute zero, he will still find a zero-point motion for his harmonic
oscillator given by $\left\langle x^{2}\right\rangle =\hbar/(2m\omega
),\left\langle \dot{x}^{2}\right\rangle =\hbar\omega/(2m),$ and%

\begin{equation}
\left\langle \mathcal{E}(\omega_{0},0)\right\rangle =\hbar\omega/2 \label{ZP}%
\end{equation}
The classical harmonic oscillator shows a random zero-point motion derived
from the random electromagnetic radiation which exists even at the zero of
temperature. \ From the point of view of classical electromagnetism, the
zero-point energy of an oscillator is not something intrinsic to the
oscillator and the state of the charged oscillator in zero-point radiation
does not involve a non-radiating situation. \ Rather in the natural classical
electromagnetic point of view, the zero-point electromagnetic radiation
(corresponding to a solution of the homogeneous Maxwell equations) produces an
oscillation of the oscillator which involves a balance between the absorption
and emission of radiation.

\subsubsection{Zero-Point Radiation vs Thermal Radiation in Classical
Electromagnetism}

It should be noted that although zero-point radiation in Eq. (\ref{ZPR2}) and
thermal radiation at $T\geq0$ appearing in Eq. (\ref{Cf}) are treated on the
same footing in classical electromagnetism, they have different effects
because of the differences in the electromagnetic spectra at zero temperature
and at finite temperature. \ The spectrum of zero-point radiation (\ref{ZPR2})
holding at zero temperature is Lorentz invariant, scale invariant, and
invariant under adiabatic compression.\cite{B1975} The zero-point radiation
spectrum is isotropic in any inertial frame.\ In free space, the zero-point
spectrum can not give rise to velocity-dependent forces on particles because
of its Lorentz-invariant character. \ The correlation functions involving
zero-point radiation depend upon only the geodesic separation between the
spacetime points at which it is evaluated.\cite{B2013} \ In contrast, the
spectrum of thermal radiation (\ref{Cf}) (including the zero-point radiation)
at a non-zero temperature $T>0$ has a preferred reference frame, that unique
frame in which the spectrum is isotropic. \ At finite non-zero temperature in
free space, the thermal spectrum gives velocity-dependent forces on any system
having electromagnetic interactions which is moving through the thermal
radiation. \ The differences in the spectra provide a qualitative classical
way for thinking about many aspects of thermal versus zero-point energy.

\section{Discussion}

\subsubsection{Connections Between Classical Electromagnetism and Modern
Physics}

The description of classical electromagnetic physics which we have given here
is fully in accord with classical electromagnetic theory and with experiment.
However, it also shows a direct connection with certain basic aspects of a
course in modern physics. \ Harmonic oscillators are treated not only in
electromagnetism texts but also in texts of modern physics and of quantum
mechanics, where a quantum oscillator has a zero-point energy in its ground
state. If the quantum harmonic oscillator is charged, then the oscillator
ground state is said to have zero-point energy but involves a non-radiating
state. \ Although quantum physics and classical electromagnetism describe
harmonic oscillators differently, they both arrive at the same results for the
mean-square displacement, mean-square velocity, and average energy. \ Indeed,
one can prove a general connection between the results of quantum theory and
the results of classical electrodynamics with classical electromagnetic
zero-point radiation for free fields and harmonic oscillator
systems.\cite{B1975b} The classical electromagnetic theory including classical
zero-point radiation can give descriptions of Casimir forces, van der Walls
forces, harmonic oscillator systems, specific heats of solids, diamagnetism,
blackbody radiation, and the absence of atomic collapse in
hydrogen.\cite{review} The theory gives a satisfying classical perspective on
a number of phenomena. However, it should be emphasized that the classical
electromagnetic theory here is not equivalent to quantum theory and is not a
substitute for quantum theory. We still do not know the areas of agreement and
disagreement between nature and classical electrodynamics including classical
zero-point radiation. \ Our ignorance regarding the results of the classical
theory compared to quantum theory is perhaps not surprising. \ It should noted
that quantum theory has been developed for over a century by a vast number of
physicists. The inclusion of classical zero-point radiation within classical
electrodynamics has been explored systematically only within the last
half-century by a tiny group of physicists.

\subsubsection{Treatment of Thermal Radiation in Classical and Quantum
Physics}

It is noteworthy that quantum physics makes a clear distinction between
zero-point energy and the thermal energy of excited states, whereas classical
electrodynamics regards both these energies in Eq. (\ref{Cf}) as having the
same character. \ Quantum ideas first arose in an attempt to interpret the
experimental data on blackbody radiation at the turn of the 20th century, and,
on this account, quantum theory makes a distinction between the thermal
\textquotedblleft photons\textquotedblright\ which give the Planck spectrum in
Eq. (\ref{LP}) and the zero-point energy which involves no photons. \ On the
other hand, classical electrodynamics with classical electromagnetic
zero-point radiation arose largely from attempts to understand Casimir forces
and has no basis for any such distinction between zero-point energy and
thermal energy . \ The experimental data for Casimir forces requires that the
full spectrum of random classical radiation corresponds to Eq. (\ref{Cf}),
including both zero-point radiation and the thermal contributions. \ Thus
within classical electrodynamics, zero-point radiation is a continuous part of
the random radiation spectrum which includes thermal radiation. \ Indeed, the
treatment of thermal fluctuations and the calculations of Casimir forces and
of van der Waals forces illustrate the distinctions between the classical and
quantum points of view.\cite{review} \ Within quantum theory, thermal
fluctuations involve the wave-particle duality of nature, whereas within
classical electrodynamics, thermal fluctuations involve simply the
fluctuations of the full random electromagnetic field above the zero-point
background. \ Within quantum physics, Casimir forces and van der Waals forces
are calculated using both the ground state zero-point energy and then separate
contributions from the thermal photons responsible for the spectrum in Eq.
(\ref{LP}). \ Within classical electrodynamics, Casimir forces and van der
Waals forces are calculated once using the full spectrum of random classical
electromagnetic given in Eq. (\ref{Cf}). \ For these force calculations, the
classical electromagnetic calculations are distinctly easier yet arrive at
exactly the same results as the quantum calculations.\cite{review}

\subsubsection{Does Classical Electromagnetic Zero-Point Radiation Exist?}

Some physicists are strongly antagonistic to the idea of \textquotedblleft
classical\textquotedblright\ zero-point energy; they insist the classical
zero-point energy does not exist and that Planck's constant is a
\textquotedblleft quantum constant.\textquotedblright\ \ Does classical
zero-point energy actually exist? \ The answer is analogous to that as to
whether or not gravitational forces exist. \ In newtonian physics, the planets
follow orbits around the sun due to gravitational forces. \ Thus in the
context of newtonian physics, we would say that gravitational forces certainly
exist. \ However, within general relativity, planets follow their paths around
the sun due to the curvature of spacetime, and we do not speak of the
existence of gravitational forces because their role has been replace by that
of curved space . \ Both quantum physics and classical physics allow the idea
of zero-point energy. \ If one wishes to describe nature in terms of classical
physics, then classical zero-point energy certainly exists, arising from the
classical electromagnetic zero-point radiation which is the boundary condition
on the homogeneous Maxwell equations, as described in the present article.
\ In quantum physics, the corresponding role is played by \textquotedblleft
quantum zero-point energy.\textquotedblright\ \ 

The objection that Planck$^{\text{'}}$s constant $\hbar$ is a
\textquotedblleft quantum constant\textquotedblright\ and hence is not allowed
in classical theory represents a misunderstanding of the nature of fundamental
physical constants. \ Planck's constant $h=2\pi\hbar$ was introduced into
physics in 1899 as a parameter in the fit to the experimental data on
blackbody radiation.\cite{DSB} \ Thus the constant $\hbar$ (or $h)$ is a
fundamental constant of nature which need not have a connection to any
particular theory. \ Within quantum physics, $\hbar$ sets the scale for the
quantum of action. \ Within classical electrodynamics, $\hbar$ sets the scale
of random classical radiation needed to describe Casimir forces, and so sets
the scale of classical electromagnetic zero-point radiation; the zero-point
energy of the radiation reappears as zero-point energy for all systems which
have electromagnetic interactions.

\end{document}